\documentclass[mathleft]{an}
\usepackage{graphicx}
\usepackage{times}
\usepackage{natbib}
\bibpunct{(}{)}{;}{a}{}{,}

\renewcommand{\vec}[1]{\mbox{\boldmath $#1$}}

\renewcommand{\vec}[1]{\mbox{\boldmath $#1$}}
 
\def\q{\qquad} 
 
\def\beg{\begin{eqnarray}} 
\def\ende{\end{eqnarray}}

\def\i{{\rm i}}
\def\d{{\rm d}}
\def \Om  {{\it \Omega}}
\def \rin {r_{\rm in}}

\def \Rm {\ensuremath{\rm{Rm}}}

\newcommand{\gsim}{\lower.7ex\hbox{$\;\stackrel{\textstyle>}{\sim}\;$}}
\newcommand{\lsim}{\lower.7ex\hbox{$\;\stackrel{\textstyle<}{\sim}\;$}}
\renewcommand{\vec}[1]{\mbox{\boldmath $#1$}}
\def\curl{{\rm curl}}

\def\XA{{\tilde A}}
\def\XB{{\tilde B}}

\def\ara\&a{ Ann. Rev. Astronomy Astrophysics}

\begin{document}

\Pagespan{789}{}%
\Yearpublication{2015}%
\Yearsubmission{2014}%
\Month{01}%
\Volume{336}%
\Issue{88}%

\title{On the toroidal-velocity antidynamo theorem under the presence of non-uniform 
electric conductivity}
\titlerunning{The toroidal antidynamo theorem   with non-uniform
conductivity}

 \author
{
G. R\"udiger$^{1,2}$\thanks{E-mail: GRuediger@aip.de} \and
M. Schultz$^1$}
\institute
{$^1$Leibniz-Institut f\"ur Astrophysik Potsdam, An der Sternwarte 16, 14482
Potsdam, Germany\\
$^2$ University of Potsdam, Institute of Physics and Astronomy,
Karl-Liebknecht-Str. 24-25,
14476 Potsdam, Germany
}
 \authorrunning
{
G. R\"udiger \and
M. Schultz}
\received{01 July 2021}
\accepted{later}
\publonline{later}


\abstract{
Laminar electrically conducting Couette flows with the hydrodynamically stable quasi-Keplerian rotation profile and non-uniform  conductivity are  probed for dynamo instability. In spherical geometry the equations for the poloidal and the toroidal field components completely decouple, resulting in free   decay, regardless of the spatial distribution
of the  electric conductivity. In  cylindrical geometry  the poloidal and toroidal components do not decouple, but here also we do not find dynamo excitations for the cases that the electric conductivity only depends on the radius or -- much more complex-- that it only depends on the azimuthal or the axial coordinate. The transformation of the plane-flow dynamo model of \cite{BW92}  to cylindrical or spherical geometry therefore fails. It is also shown that  even the inclusion  of axial flows of both directions does  {\em not} support the dynamo mechanism. The Elsasser toroidal-velocity antidynamo theorem,   according to which dynamos without any radial velocity component cannot work, is  thus not softened by   non-uniform conductivity distributions.  } 
 
\maketitle

\section{Introduction}
It was demonstrated by \cite{BW92} that a large-scale magnetic field
can be excited as  a slab
dynamo formed by a plane laminar shear flow and   an electric conductivity
which is sinusoidally
modulated in the stream-wise direction. The flow becomes supercritical for
rather large magnetic
Reynolds numbers of  $O(10^{4...5})$, which can be optimized when the
perturbation scale of the
magnetic field approaches the scale of the conductivity variations, where
the latter may exceed
the thickness of the slab. If, on the other hand, the electric conductivity
varied crosswise
rather than streamwise no dynamo effect was found by the authors.

This raises the question of whether this positive result can be transformed to
cylindrical and/or
spherical geometry. If yes then a dynamo should also work for differential
rotation of a fluid
of non-uniform electric conductivity. The dynamo action should be possible
for non-rigidly
rotating cylinders or spheres with non-uniform conductivity distribution in the azimuthal direction.
It is known, however,
that spherical  dynamos with only toroidal internal flows do not exist for uniform conductivity distribution
\citep{E46,IJ88long,KB17}.

  In this
case differential rotation is not enough to form a dynamo mechanism but, as
suggested by \cite{G70},
pure meridional flow should be enough for sufficiently large  magnetic Reynolds number of the flow \citep{M06,LI10}. \cite{DJ89} demonstrated how the
addition of
differential rotation to the meridional flow can strongly reduce the
critical magnetic Reynolds
number for suitable amplitude ratios of
the flow components.

One can easily show that for spheres  the toroidal-velocity antidynamo theorem also
holds for
non-uniform conductivity distribution.
The induction equation for a laminar magnetized fluid with a finite
magnetic resistivity
$\eta=1/\mu\sigma$ (with $\sigma$ as its electric conductivity) is
\begin{eqnarray}
\frac{\partial \vec{B}}{\partial t}= {\textrm{curl}} (\vec{u} \times
\vec{B} - \eta\
\curl\vec{B})
   \label{mhd0}
\end{eqnarray}
with $ {\textrm{div}}\ \vec{u} = {\textrm{div}}\ \vec{B} = 0$ for an
incompressible fluid and 
$\vec{u}$ as its  velocity. $\vec{B}$ is  the magnetic field; here  the magnetic
resistivity is
not necessarily uniform. The question is whether for given mean flow this
equation, which is
homogeneous in the magnetic field, has eigenvalues with marginal growth
rates.

Following the method demonstrated in the textbook of \cite{KR80}, the
induction equation in spherical geometry is
split into poloidal $\curl\vec{A}_{\rm tor}$ and toroidal $\vec{B}_{\rm tor}$
components
\begin{eqnarray}
\vec{A}_{\rm tor}= - \vec{r}\times \nabla S,\ \ \ \ \ \ \ \ \ \ \ \ \ \ \ \
\ \ \ \ \ \ \ \
\vec{B}_{\rm tor}= - \vec{r}\times \nabla T,
   \label{mhd01}
\end{eqnarray}
with $\vec r$ as the radial vector ${\vec r}=(r,0,0)$ in spherical
geometry. The result is two equations for
$S$ and $T$. Here only the relation
\begin{eqnarray}
\frac{\partial S}{\partial t}+\Om \frac{\partial S}{\partial \phi}- \eta
\Delta S = U_{\rm
ind}
   \label{mhd02}
\end{eqnarray}
for the poloidal field component $S$ must be considered, where $\Om$ is the
(non-uniform)
rotation rate of the sphere. The right-hand side of this equation describes the
induction by the mean
circulation ${\vec u}^{\rm m}$, i.e.
\begin{equation}
{\cal L} U_{\rm ind} = - \vec{r}\cdot \curl\ {\vec u}^{\rm m}\times {\vec
B}.
   \label{mhd03}
\end{equation}
The  operators
\beg
&\Delta&= 
\frac{1}{r^2}\Big(\frac{\partial}{\partial r}(r^2\frac{\partial}{\partial
r}) +{\cal L}\Big) , \nonumber\\
&{\cal L}&= 
    \frac{1}{\sin\theta}\frac{\partial}{\partial\theta}\Big(\sin\theta
    \frac{\partial}{\partial\theta}\Big) +
    \frac{1}{\sin^2\theta}\frac{\partial^2}{\partial\phi^2}
  \label{mhd04}
\ende
are well-known. 
From Eq.  (\ref{mhd03}) one finds the explicit formulation
\begin{eqnarray}
{\cal L} U_{\rm ind}& =&
- \nabla (r u_r)\cdot {\vec r}\times \nabla T  
-\vec{u}^{\rm m} \nabla {\cal L}S+\nonumber\\
& +& u_{r} (r \Delta S+\nabla
\frac{\partial}{\partial r}(rS)). 
\label{mhd05}
\end{eqnarray}
The striking result is that only the first term of the right-hand side
 of this relation
is able to couple
the toroidal field with Eq. (\ref{mhd02}), but this term vanishes for
$u_r=0$. Without this
source term the relation (\ref{mhd02}) is homogeneous in $S$ and only
describes the decay of
this
field component, regardless
 of the actual form of the function $\eta$  \citep{IJ88}. A
self-maintaining
nonturbulent dynamo in spherical geometry is thus only possible for 
flows  with
finite  radial component, i.e. including meridional flows.
This is the well-known
antidynamo theorem, which obviously also holds for non-uniform distributions
with $\eta>0$. The
transformation to spherical geometry of the plane dynamo model of \cite{BW92}
that relies on streamwise
$\eta$-modulation between two narrow plates is thus not possible. \cite{KB17} have shown that this toroidal-velocity antidynamo theorem also survives up to    a certain  nonradial  resistivity variation  which decreases for  increasing  magnetic Reynolds numbers. The existence of this   limit  characterizes  the robustness of the  antidynamo theorem; dynamo excitation, therefore, is not excluded by this analysis for sufficiently large resistivity variations. 

This might support the idea of \cite{RM17} that a dynamo can basically work
in the thin stably
stratified atmosphere of a hot Jupiter on the basis of strong conductivity
variations due to
the heating of the near host star. According to the findings of the present paper, however, even this
 dynamo can only
operate under the presence of differential rotation {\em and} meridional
circulation. Detailed
calculations of the wind system in ultra-hot Jupiters have been published
by
\cite{TK19}.

For non-uniform magnetic resistivity the induction equation reads
\begin{eqnarray}
\frac{\partial \vec{B}}{\partial t}= {\textrm{curl}} (\vec{u} \times
\vec{B}) + \eta
\Delta\vec{B} - \nabla\eta \times \curl \vec{B}
   \label{mhd1}
\end{eqnarray}
\citep{GN10,RM17}. It has been argued that the last term   of this equation can  formally  be considered as a new alpha-effect term allowing 
for dynamo operation. However, this term  is  
perpendicular to the electric
current $\curl\vec B$, while the corresponding term by an alpha effect is
parallel to $\curl\vec
B$. One can show that one of the main consequences of this last term is the
advection of magnetic
field in the direction of $\nabla\eta$. It is
\begin{eqnarray}
\nabla\eta \times \curl \vec{B}= \curl(\nabla \eta\times\vec{B}) +\nabla
(\nabla\eta\cdot\vec{B}) + ....,
   \label{mhd06}
\end{eqnarray}
where the dots represent a tensor formed with second derivatives of $\eta$,
i.e.
$(\Delta\eta\delta_{ij}-2\eta_{,ij})B_j$. The first term on the right-hand side of this
equation describes
advection of the field antiparallel to $\nabla\eta$ (see Eq. (\ref{mhd0})), which is, if too strong, known to suppress dynamo action, see \cite{KR80},  \cite{BM92}  and \cite{GR22}. The
remaining terms also
cannot play the role of an alpha effect.
The existence of an alpha effect always requires the existence of a
pseudo-scalar or a
pseudo-tensor.

It makes sense  to inspect the radial component of (\ref{mhd1}) in spherical coordinates, i.e.
\begin{eqnarray}
\frac{\partial {B_r}}{\partial t}&+&\Om\frac{\partial B_r}{\partial\phi}=
\eta \big(
\frac{1}{r}\frac{\partial^2}{\partial r^2}r B_r
+ \frac{1}{r^2}\frac{\partial^2 B_r}{\partial\theta^2}+\nonumber\\
&+&
 \frac{1}{r^2\sin^2\theta}\frac{\partial^2 B_r}{\partial\phi^2}
+ \frac{\cot\theta}{r^2}\frac{\partial B_r}{\partial\theta}+
\frac{2}{r}\frac{\partial B_r}{\partial r}+\frac{2 B_r}{r^2} \big)-\nonumber\\ 
&-&
\frac{1}{r^2}\frac{\partial\eta}{\partial
\theta}\big(\frac{\partial}{\partial r}r B_\theta
-\frac{\partial B_r}{\partial \theta}\big)+\nonumber\\ 
&+&\frac{1}{r^2\sin\theta}\frac{\partial
\eta}{\partial\phi}\big(\frac{1}{\sin\theta}\frac{\partial
B_r}{\partial\phi}-\frac{\partial}{\partial r}r B_\phi
  \big) 
  \label{mhd2}
\end{eqnarray}
All terms due to a meridional circulation have been cancelled. We note 
for the last two  lines of
this expression that  the $\nabla$-terms of (\ref{mhd06}) appear. For strong shear the
combination of
$\partial \eta/\partial \phi$ with $B_\phi$ should make the strongest
effect. A radial gradient
of the resistivity does not appear explicitly.
\section{Cylindrical geometry}
The  negative conclusion for spherical geometry  cannot automatically be transformed to
cylindrical geometry. A
special situation exists for cylindrical geometry as the operator
{\em curl curl} combines various
magnetic field components, i.e.
\begin{eqnarray}
\frac{\partial {B_R}}{\partial t}+\Om\frac{\partial B_R}{\partial\phi}&=&
\eta \big(
\frac{\partial}{\partial R}(\frac{1}{R} \frac{\partial }{\partial R}(R B_R) )
+ \frac{\partial^2 B_R}{\partial z^2}+\nonumber\\
&+& \frac{1}{R^2}\frac{\partial^2 B_R}{\partial\phi^2}
-\frac{2}{R^2}\frac{\partial B_\phi}{\partial \phi} \big)
+....,
  \label{mhd3}
\end{eqnarray}
now written in cylindrical coordinates $(R,\phi,z)$. Obviously, in contrast to (\ref{mhd2}), for nonaxisymmetric fields a
coupling exists between the
azimuthal and the radial field components which might be able to create a
dynamo
in a similar sense as the $\vec{\Om}\times \vec{J}$ effect in turbulent
dynamo theory does. The
latter is part of the eddy diffusivity tensor containing the
linear-in-$\Om$ influence on the
magnetic dissipation. It is thus necessary to find out whether for
cylindrical models the
combination of differential rotation and non-uniform electric conductivity
can lead to dynamo instability
\subsection{The model}
Distances and wave numbers may be normalized with the outer cylinder
radius
$R_0$, and the inner cylinder is $\rin R_0$. The rotation and other
frequencies are
normalized with the rotation rate $\Om_{\rm in}$ of the inner cylinder; the
rotation rate of
the outer cylinder is $\mu \Om_{\rm in}$. The rotation law may be written
as
\beg
\Om(R)=a_\Om+\frac{b_\Om}{R^2}
\label{Omeg}
\ende
with
\beg
a_\Om= \frac{\mu-\rin^2}{1-\rin^2},
\q\q\ \ \ \ \ b_\Om=
\frac{1-\mu}{1-\rin^2}\rin^2.
\ende
Here $\mu=\Om_{\rm out}/\Om_{\rm in}$ is the ratio of the rotation rates of the outer and inner cylinders. For $\mu=\rin^{1.5}$ the cylinders rotate according to the Kepler law, i.e. $\mu=0.35$ for $\rin=0.5$. This rotation law is flat enough to avoid  the flow becoming centrifugally unstable. The Rayleigh limit for $\rin=0.5$ is $\mu=0.25$, above which the flows are stable.
The magnetic Reynolds number of the rotation is defined as 
\beg 
{\Rm}= \frac{{\Om_{\rm in}} R_0^2}{\eta_0},
\ende
 where $\eta_0$ is a characteristic value of the resistivity defined by the definition of the $\eta$-profile. The magnetic field vector in  cylindrical coordinates is $\vec{B}=(A,B,B_z)$.
The equations are solved by means of the Fourier series $A=\sum_m A_m(R)
\exp{\i{(kz+m\phi+\omega t)}}$ and $B=\sum_m B_m(R) \exp{\i{(kz+m\phi
+\omega t)}}$, while the
axial field $B_z$ is always substituted by means of the  divergency condition. The solutions are assumed to be periodic  along the rotation axis, with $k$
as the  wave
number (a parameter) in this direction. $\omega$ is the eigenvalue of the system, with its real part as the azimuthal drift rate and its negative
imaginary part as the
growth rate. Negative growth rate indicates decay of the magnetic pattern while vanishing growth rate marks neutral instability. In case of self-excitation the wave number $k$ can be used to find the optimal excitation condition.

The eigenfunctions and the eigenvalues fulfil the symmetry conditions
\beg
A_{-m}=A_m^*, \ \ \ \ \ \ B_{-m}=B_m^*\ \ \ \ \ \ \ \ {\rm if}\ \ \
\omega_{-m}=-\omega^*_m,
\ende
where an asterisk represents the conjugate complex expression.

To normalize the growth rate with the diffusion time we have simply to write
\beg
\omega_{\rm gr}=- \Rm\ \omega^{\rm I}.
\label{omgr}
\ende
Here the notation is  $\omega=\omega^{\rm R}+\i\omega^{\rm I}$. If $\omega_{\rm gr}$ does not depend on $\Rm$ and is of order unity then growth (or decay) of the modes are only directed by the
diffusion scale.
Negative values of (\ref{omgr}) indicate decaying magnetic fields. All plots in this paper
are using the growth rates described in these units. The same is true for the drift rates $\d \phi/\d t = -\omega_{\rm dr}$ with
\beg
\omega_{\rm dr}=  \Rm\ \omega^{\rm R}.
\label{omdr}
\ende
Hence,  $\omega_{\rm dr}<0$ implies azimuthal drift in the positive direction.

\begin{figure}
  \centerline{
  \vbox{
  \includegraphics[width=0.47\textwidth]{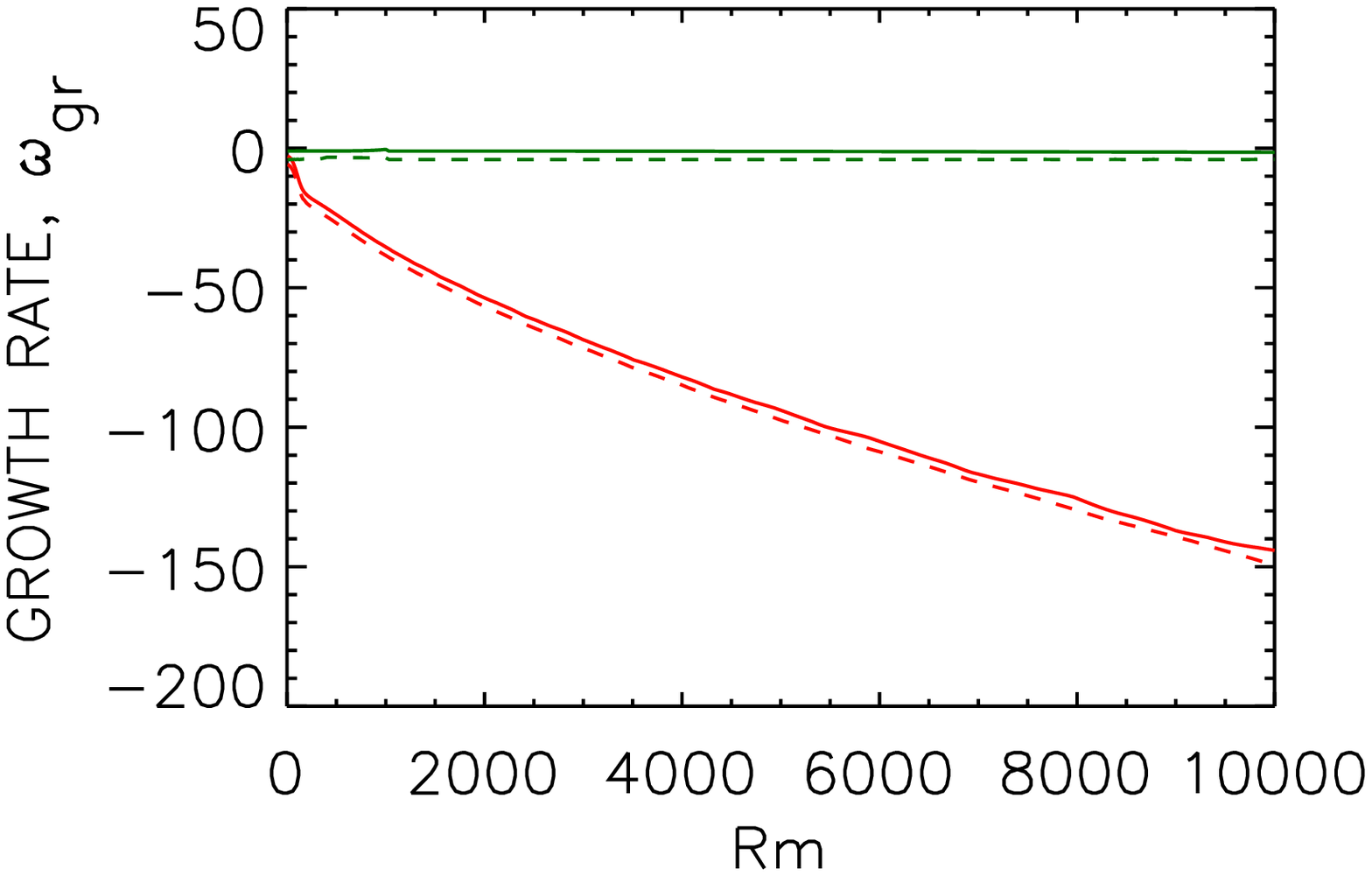}
  \includegraphics[width=0.47\textwidth]{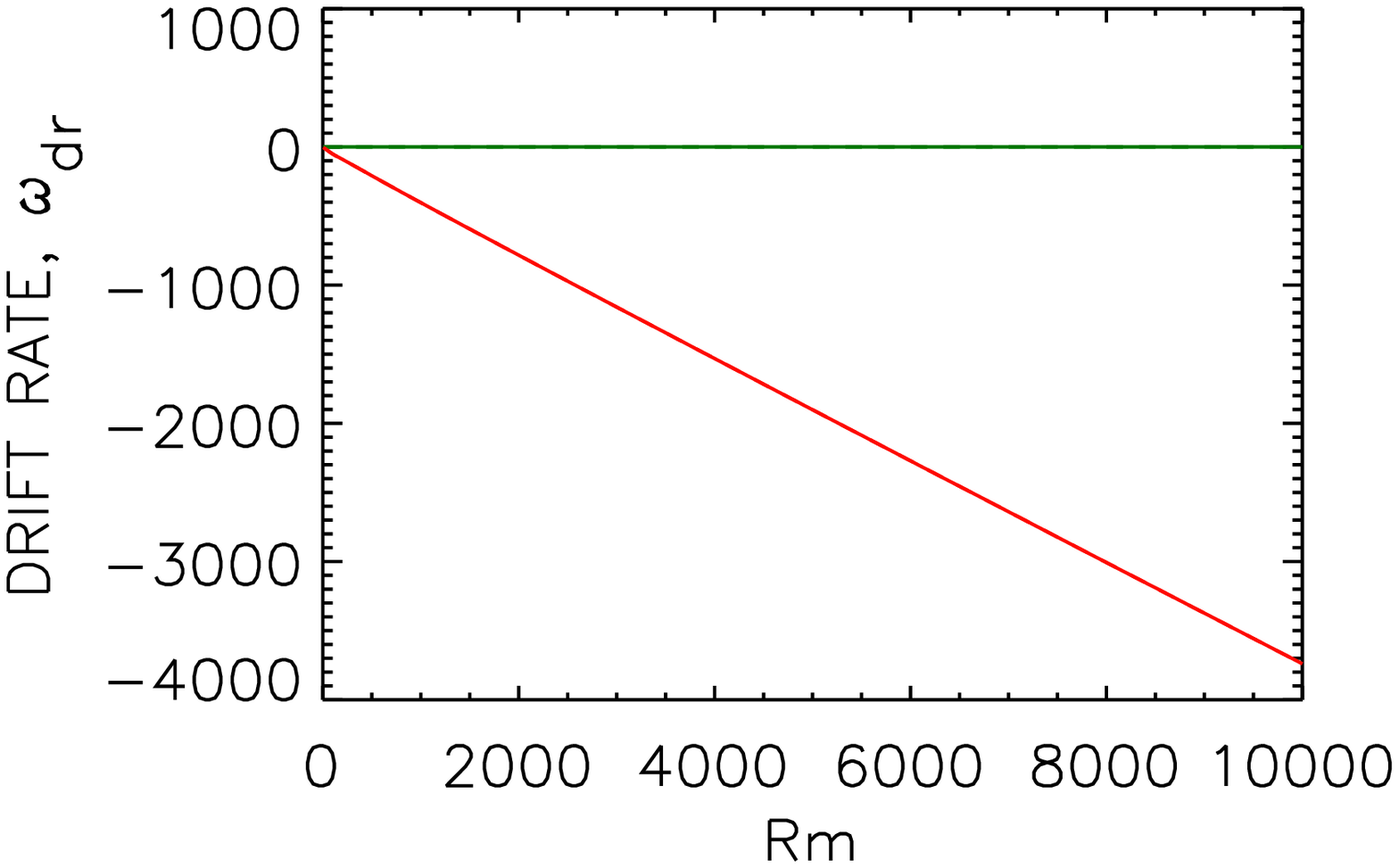}}}
\caption{Reference plots:  decay rates (top) and drift rates (bottom) in units
of the diffusion
time for uniform $\eta$ and for $m=0$ (green) and for  $m=1$ (red). Two different wave numbers
$k=1$ (solid), $k=2$ (dashed). $\rin=0.5$,
$\mu=0.35$ (Keplerian rotation), and perfectly conducting boundary conditions.
 }
\label{refer}
\end{figure}

For comparison both the
decay  and drift rates of infinite cylinders 
with uniform resistivity have been calculated with the results presented by
Fig. \ref{refer}.
The upper panel concerns the decay of the modes with $m=0$ (green lines) and $m=1$ (red lines) for two different wave numbers $k=1$ (solid lines) and $k=2$ (dashed lines). Axisymmetric  magnetic fields decay with the diffusion time
scale, hence  the
influence of the differential rotation 
can be neglected.  The nonaxisymmetric mode with $m=1$ 
decays even faster. Without rotation the resulting  growth rates for $k=1$ are 
 $\omega_{\rm gr}=-1$ for axisymmetric fields and  $\omega_{\rm gr}=-2.83$ for nonaxisymmetric fields with $m=1$. For $k=2$ the corresponding numbers are  $\omega_{\rm gr}=-4$ and  $\omega_{\rm gr}=-5.83$.
In
contrast to the decay law
for $m=0$, for $m=1$ the influence of the magnetic Reynolds number is strong so
that  the
physical decay rate scales with the rotation rate, $\omega_{\rm gr}\propto
-\Om$. The decay time of
nonaxisymmetric fields is strongly reduced by rotational shear, independent
of its sign. As the winding-up of the field lines by the differential rotation
leads to shorter scales, for rapid rotation the decay time becomes shorter \citep{KR80}. For high magnetic Reynolds numbers nonaxisymmetric fields decay about $\sqrt{\Rm}$ times faster than axisymmetric fields.  

The azimuthal drift of an axisymmetric field, of course, vanishes 
while  nonaxisymmetric fields  basically drift by the action of the
differential rotation
(bottom panel of Fig \ref{refer}). The drift frequency for large $\Rm$ corresponds to the rotation rate of the outer cylinder, i.e. $\mu\Om_{\rm in}$. As formulated  by \cite{E46}, assuming uniform conductivity  ``for toroidal flow the induction effect consists in
oscillatory changes of the field amplitudes superposed
upon the slow, general decay of the field''.

In the following  we shall demonstrate that these results are only
slightly modified if
the molecular resistivity $\eta$ is no longer uniform. Here we distinguish between models with radial gradients and models with azimuthal gradients. Axial gradients of $\eta$ are not considered in this paper as \cite{BW92} also did not find dynamo excitation for such models.
\subsection{Radius-dependent conductivity}
We start by considering a differentially rotating  fluid in a cylindrical container  with a radius-dependent
magnetic resistivity.
We write
$
\eta=\eta_0 \eta(R)$ with  $\eta(R)=R^{-\mu_\eta}$ (hence $\eta(1)=1$),
where the profile parameter $\mu_\eta$ can have both signs. For positive
values the electric
conductivity grows outward. The model is rather 
simple as the
radius-dependent magnetic resistivity does not
lead to complicated mode couplings as in the model by \cite{BW92} or as in  our
 models with
azimuth-dependent or $z$-dependent $\eta$ profiles.
The equation system for four differential equations of first order is used as
formulated by
\cite{SR02}, i.e.
\beg
\frac{\d \XA_m}{\d R} -( k^2+\frac{m^2}{R^2}) A_m &-&\frac{\i \Rm}{\eta(R)}
(\omega+m\Om(R))A_m-\nonumber\\
&-&\frac{2\i m}{R^2}B_m= 0,
\label{g2}
\ende
\beg
\frac{\d \XB_m}{\d R} &-&\frac{\mu_\eta \XB_m}{R} -(k^2+\frac{m^2}{R^2}) B_m-\nonumber\\
&-&\frac{\i\Rm}{\eta(R)} (\omega+m\Om(R))B_m+\nonumber\\
&+&(2+\mu_\eta)\frac{\i m A_m}{R^2} -2\frac{{\Rm}\ b_\Om}{\eta(R) R^2}A_m
= 0
\label{g1}
\ende
with 
\beg
\XA_m=\frac{\d A_m}{\d R} + \frac{A_m}{R}, \ \ \ \ \ \ \ \ \ \ \ \ \ \ \ \
\ \
\XB_m=\frac{\d B_m}{\d R} + \frac{B_m}{R}.
\label{g33}
\ende
The boundary conditions for these equations are 
$
A_m=\XB_m =0 
$
 for perfectly conducting walls at $R=\rin, 1$ or  
\beg
\left( k^2 +\frac{m^2}{R^2} \right) B_m -\frac{\i m   \XA_m}{R} =0
\label{eq8}
\ende
for vacuum conditions.  In this case it is also
\beg
k A_m-\left(\XA_m+\frac{\i m B_m}{R}\right)\left(\frac{m}{kR} +\frac{
I_{m+1}(kR)}{I_m(kR)}
\right)=0
\label{eq9}
\ende
at $R=\rin$  and 
\beg
k A_m-\left(\XA_m+\frac{\i m B_m}{R}\right)\left(\frac{m}{kR} -\frac{
K_{m+1}(kR)}{K_m(kR)}
\right)=0
\label{eq10}
\ende
at $R=1$ \citep{RGH18}. We note that the modified Bessel functions $I$ and $K$ satisfy  $I_{-m}=I_m$ and  $K_{-m}=K_m$.
Sometime also pseudo-vacuum conditions (vertical fields)
$B_m=\XA_m =0 $
have been used at  $R=\rin, 1$. 

\begin{figure}
  \centerline{
  \vbox{
  \includegraphics[width=0.47\textwidth]{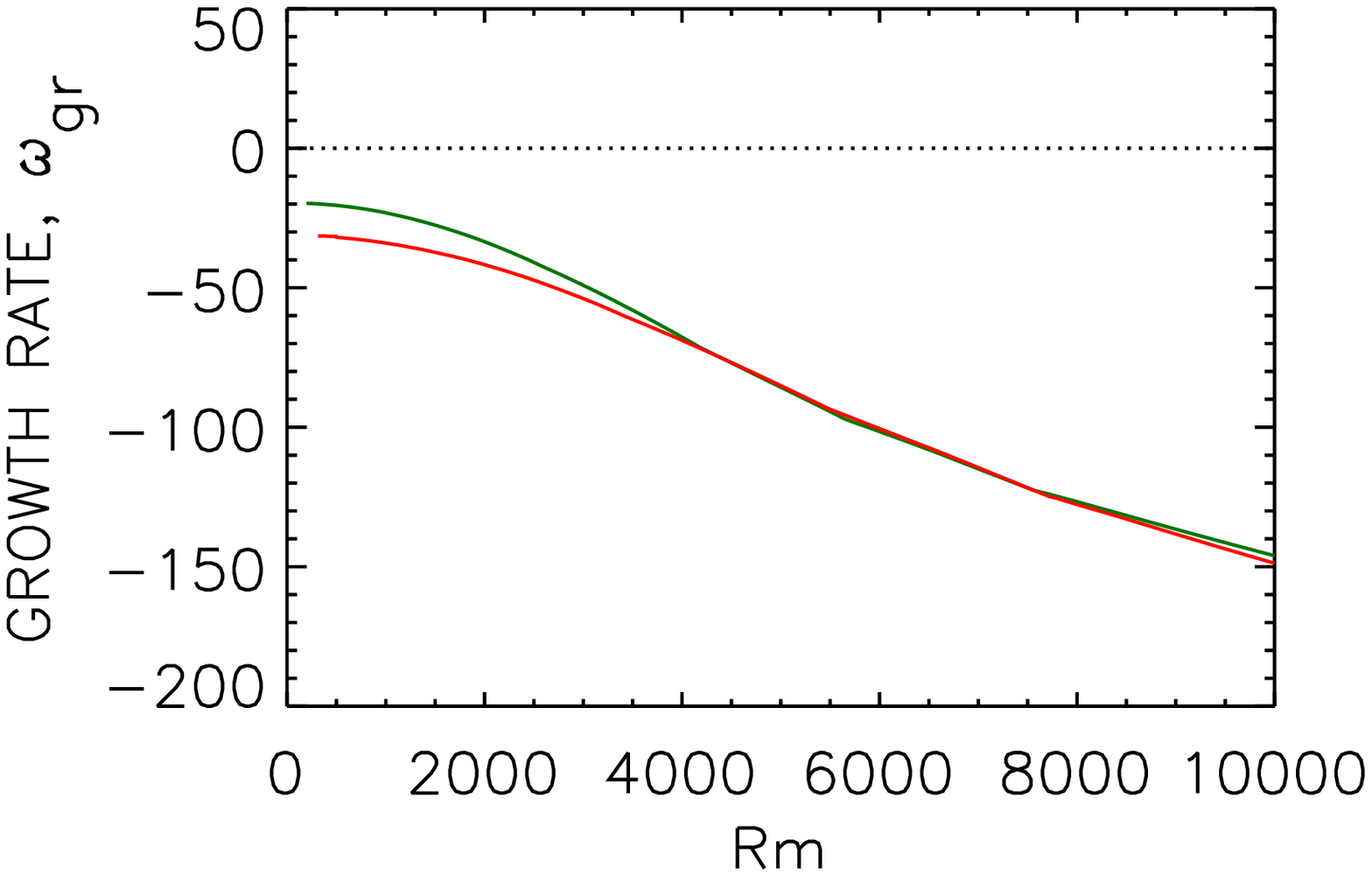}
  \includegraphics[width=0.47\textwidth]{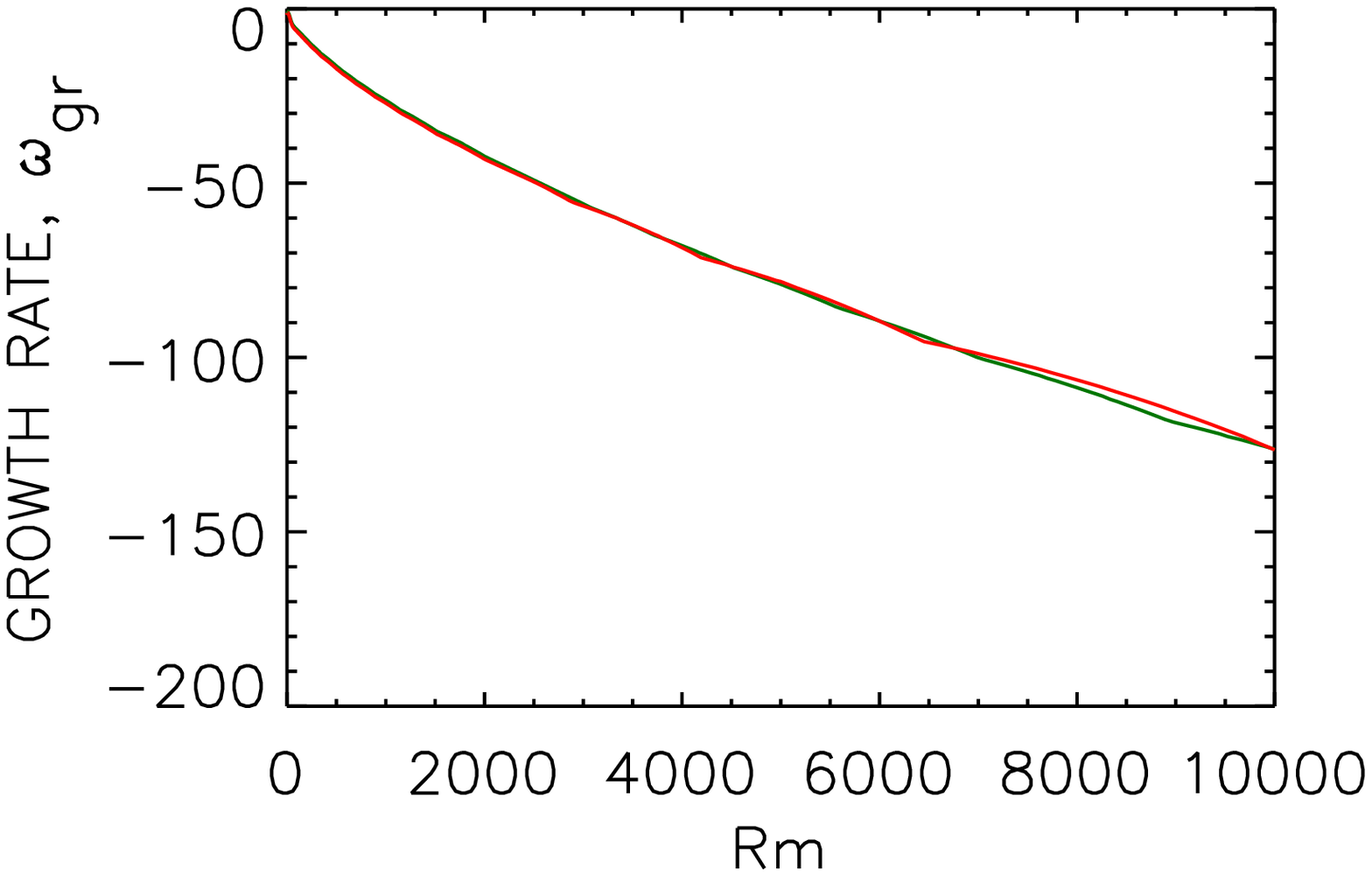}}}
\caption{Decay rates of the nonaxisymmetric mode ($m=1$) in units of the diffusion time for two models of
$\eta=\eta(R)$. Top:
$\mu_\eta=10$ (conductivity grows outward), insulating  walls; Bottom: $\mu_\eta=-3$ (conductivity grows inward), perfectly conducting walls. Wave numbers $k=0.1$ (green), $k=1$
(red).
$\rin=0.5$, $\mu=0.35$ (Keplerian rotation). 
}
\label{fig01}
\end{figure}

The decay rates for two models with Keplerian rotation and with radial $\eta$-gradients of
opposite signs are given in Fig. \ref{fig01}. The free axial wave numbers
have been varied
between $k=0.1$ and $k=1$, but the curves are almost identical. Magnetic
modes with different
wave numbers decay with the same rate -- except for $\Rm=0$ where the modes with 
larger wave number (here $k=1$) decay faster. The magnetic Reynolds number of the
rotation is assumed to increase up to the very large values $\Rm\lsim 10^4$. As expected for 
nonaxisymmetric fields under the influence of differential rotation, 
the decay rate grows with growing magnetic Reynolds numbers;  the
decay frequency is thus
proportional to the rotation rate hence $\tau_{\rm dec}\propto \tau_{\rm
in}$ with $\tau_{\rm
in}$ as the rotation period of the inner cylinder. Short time scales are produced by the winding-up of the field lines by the non-uniform rotation, which decay faster the larger the value of $\Rm$ is.
The decay rates for $\eta$
increasing outwards
or decreasing outwards do not differ basically. The influence
of the sign of the $\eta$-gradient on the decay times is obviously   weak.  As it should, the decay is (slightly)
faster for negative
gradients, i.e. for $\mu_\eta>0$. The drift rates (not shown) in both cases are almost identical with the values shown in the lower panel of Fig. \ref{refer}.  Independently of the
basically different radial $\eta$-profiles the nonaxisymmetric field pattern always drifts with the rate $\Om_{\rm out}$  for large $\Rm$. 

The result is that a  radial  non-uniformity of conductivity even for large magnetic
Reynolds numbers does not lead to dynamo action \citep{IJ88}. Figure \ref{fig01}  shows that 
a nonaxisymmetric  field also cannot be maintained by differential rotation alone.
The toroidal-velocity 
antidynamo theorem obviously also holds for cylindrical geometry  and arbitrary radial $\eta$-gradients. 
The curves do not show
any indication that
the trends are changed   for higher magnetic
Reynolds numbers. In agreement with \cite{KB17} we do not find  evidence for dynamo action of differential rotation in combination with  radial $\eta$-gradients of either sign.

\section{Azimuth-dependent conductivity}
Models with nonradial variation of the $\eta$-profile are more complex. The reason is that in the nonradial coordinates $z$ and $\phi$ the solutions are developed into Fourier modes, and for resistivities  varying in these directions the modes couple, always producing an infinite series. Here we shall  only provide  solutions of the dynamo equation  (\ref{mhd0}) for $\eta$ as a function of the azimuth $\phi$.

The azimuthal variation of the magnetic resistivity may be written as
\beg
\eta=\eta_0(1-\kappa \cos\phi),
\label{etaf}
\ende
where the coefficient $\kappa$ may vary between 0 and 1. The value $\eta_0$ is the
 average over the
azimuth.  The ratio of the maximum value of $\eta$ and its minimum  is simply $(1+\kappa)/(1-\kappa)$, so that it approaches large values  for $\kappa\to 1$. 

\begin{table}
\caption{Coefficients $\kappa_n$  and $\bar\kappa_n$ for
$\kappa=0.5$ and $\kappa=0.9$.
}
\label{tab4}
\centering
\begin{tabular}{l|cc|cc}
\hline\hline

\\

mode $n$&$\kappa_n(0.5)$ &$\bar\kappa_n(0.5)$& $\kappa_n(0.9)$ & $\bar\kappa_n(0.9)$ \\ \\
\hline
-3 &0.019 &0.022& 0.25 &0.56 \\ 
-2 &0.072 &0.083& 0.39 &0.96 \\
-1& 0.27 &0.31&0.63&1.44\\ 
0 &0 &1.15&0 & 2.29  \\
 1& -0.27&0.31&- 0.63 &  1.44 \\ 
2 &-0.072&0.083&-0.39&0.96 \\
 3 &-0.019&0.022&-0.25 & 0.56\\
  \hline
\end{tabular}
\end{table}

Ansatz (\ref{etaf})  describes the consequences of a heating of the cylinder at a phase zero and a cooling at the opposite phase $\pi$. One could also work with higher azimuthal frequencies by replacing $\cos \phi$ in (\ref{etaf}) by $\cos N\phi$, but here we always take $N=1$. In all cases with $\kappa\neq 0$ the magnetic modes with the azimuthal wave number $m$
are coupled, so that an infinite set of equations results. The
coupling
coefficients are
\beg
\kappa_n=\frac{\i \kappa}{2\pi} \int^{2\pi}_0\frac{\sin\phi\ e^{\i
n\phi}}{1-\kappa \cos\phi}
\d\phi, \nonumber\\
 \bar\kappa_n=\frac{
1}{2\pi}
\int^{2\pi}_0\frac{ e^{\i n\phi}}{1-\kappa \cos\phi}\d\phi.
\label{kappas}
\ende
\noindent
The $\kappa_n$ and $\bar\kappa_n$ are real numbers with $\kappa_{-n}=-\kappa_n$ and
$\bar\kappa_{-n}=\bar\kappa_n$. For $\kappa=0$ we have
$\bar\kappa_0=1$ and all other  $\bar\kappa_n$ vanish.  Analytically, for $\kappa\gg1$ it is $\kappa_{\pm1}\simeq \mp \kappa/2$ and $\bar \kappa_{\pm1}\simeq  \kappa/2$

\begin{figure}
  \vbox{
  \includegraphics[width=0.47\textwidth]{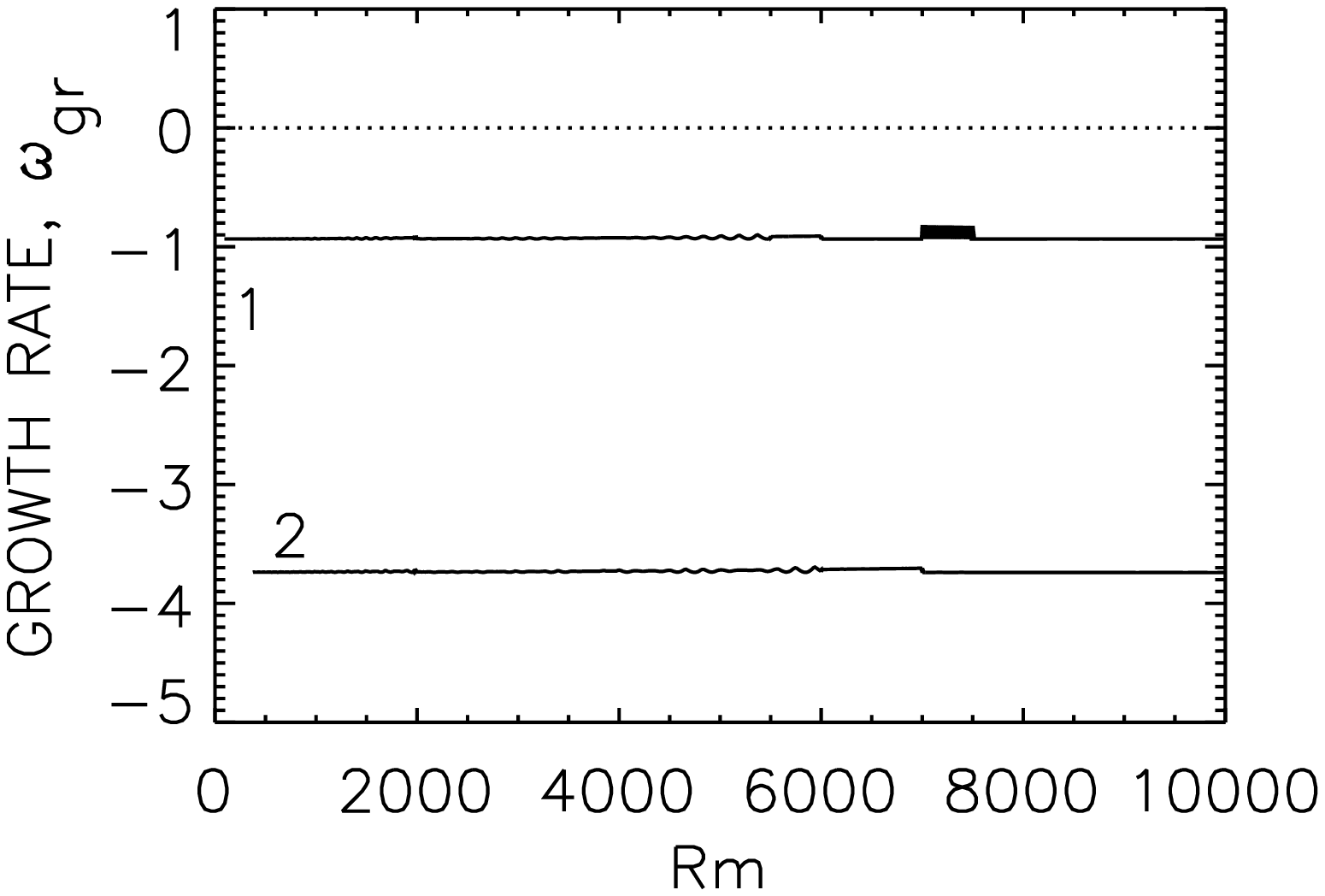}
  \includegraphics[width=0.47\textwidth]{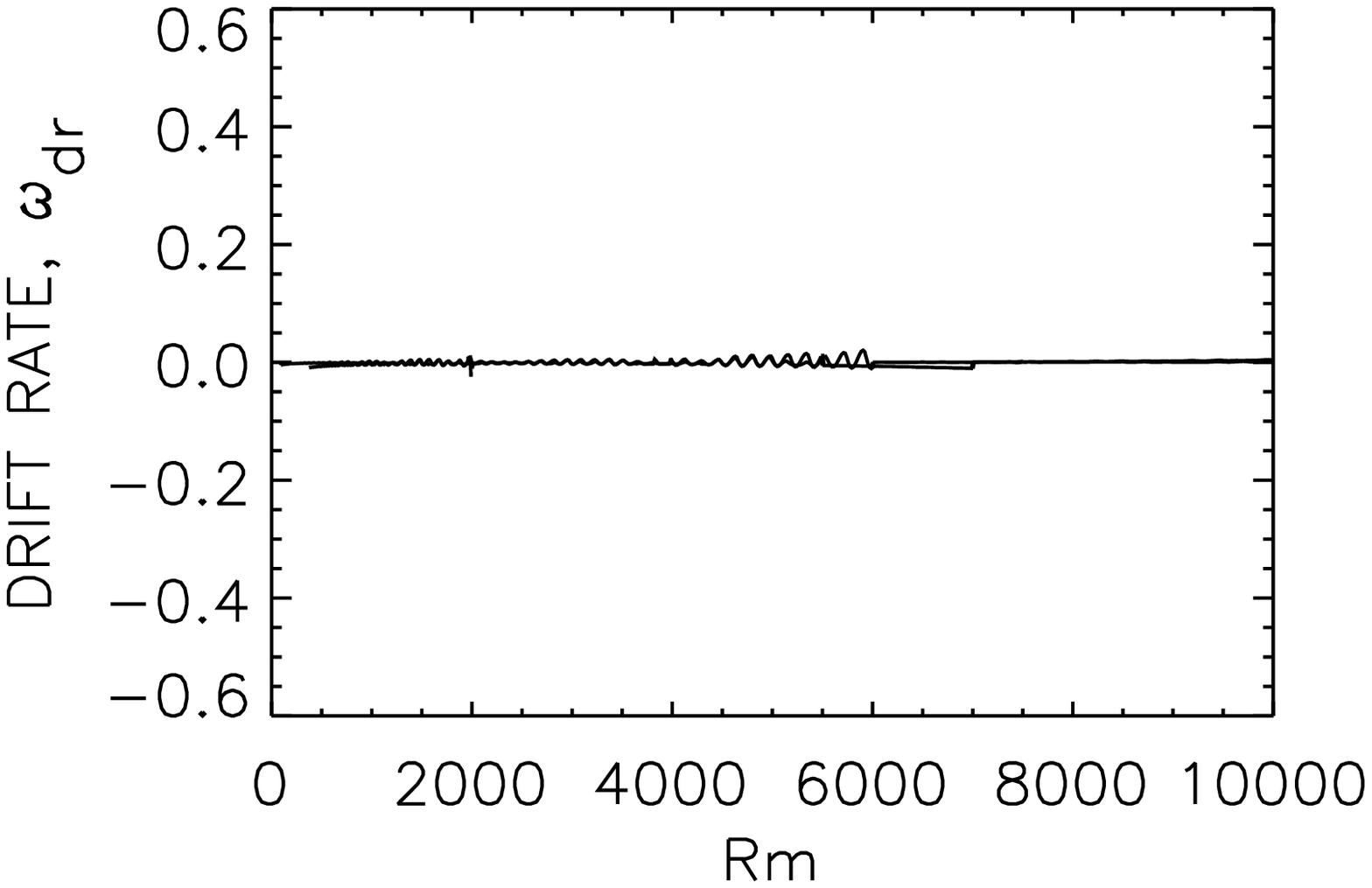}}
\caption{Growth rate (top) and drift rate (bottom)  in units of the
diffusion rate 
for two values of the free axial wave number ($k=1$ and $k=2$, marked) for
azimuth-dependent electric
conductivity. $\rin=0.5$, $\mu=0.35$ (Keplerian rotation). $\kappa=0.5$, $m=1$,
$M=1$. The dotted line denotes marginal stability. Perfectly conducting
boundary conditions.
 }
\label{fig10}
\end{figure}

According to Cowling's theorem we must consider the  growth or decay of a nonaxisymmetric magnetic field. Written with the definitions (\ref{g33}) the resulting equation system for growth or decay  of the mode with $m=1$ (defined by $M=0$)
is
\beg
\frac{\d \XA_m}{\d R}  &-&( k^2+\frac{m^2}{R^2}) A_m  -\frac{2\i m}{R^2}B_m-\nonumber\\
&-&\i {\Rm}\sum_{n=1-M}^{1+M} \bar\kappa_{n-m} (\omega+n\Om(R)+k U(R))  A_n+\nonumber\\
&+& \sum_{n=1-M}^{1+M} \kappa_{n-m} \left(\frac{ nA_n}{R^2}+\frac{\i
\XB_n}{R}\right)
= 0
\label{g5}
\ende
and
\beg
\frac{\d \XB_m}{\d R}  &-&(k^2+\frac{m^2}{R^2}) B_m+\frac{2\i m}{R^2}A_m-\nonumber\\
 & -&{\i \Rm}\sum_{n=1-M}^{1+M} \bar\kappa_{n-m}(\omega+n\Om(R)+k U(R))B_n-\nonumber\\
 &-&2{\Rm}\ \frac{b_\Om}{ R^2} \sum_{n=1-M}^{1+M}\bar\kappa_{n-m}A_n
= 0
\label{g6}
\ende
formulated for all $m$ with $1-M\leq m\leq 1+M$. The number $M$ limits the equation
system which is solved
if the eigenvalues no longer depend on the choice of $M$. It is also possible to consider the stability of a higher nonaxisymmetric mode, but here 
we   focus on  $m=1$. The numerical results in Fig. \ref{fig10} for $\kappa=0.5$  and with the approximation $M=1$
for  growth rates and drift rates (both normalized with the diffusion frequency) of the eigensolutions are plotted for our standard container  ($\rin=0.5$) and for Keplerian rotation. The resistivity varies in accordance with (\ref{etaf}) without any  radial variation. The plot shows the eigenfrequencies  for the two wave numbers $k=1$ and $k=2$.
The results are surprising. The modes are always decaying. However,  in contrast to  the eigenvalues for isolated  nonaxisymmetric modes and axisymmetric resistivity with the characteristic $\Rm$ behavior  shown in Fig. \ref{fig01}, the resulting  growth rates in units of the diffusion rate do {\em not} depend on the magnetic Reynolds number (upper panel) and the decaying patterns do {\em not} drift in azimuthal direction (lower panel). The nonaxisymmetric  $\eta$-profile stops the azimuthal drift of the nonaxisymmetric modes and the influence of the differential rotation on the normalized decay rate disappears. Due to the nonaxisymmetric $\eta$-profile the  $m=1$ mode is always coupled to the $m=0$ mode which does not  feel the influence  of the differential rotation (see Fig. \ref{fig01}, upper panel) and decays only slowly. By the coupling of all the azimuthal  modes the $m=1$ mode then also decays as slowly as the $m=0$ mode (see Fig. \ref{fig10}, upper panel) \footnote{This statement can be proved by solving the system (\ref{g5}) and (\ref{g6}) for $\kappa=0$ and $M=0$ which exactly reproduces the red lines in Fig. \ref{fig01} for the decay of the nonaxisymmetric $m=1$ mode with uniform resistivity.}.
\begin{figure}
  \centerline{
  \includegraphics[width=0.47\textwidth]{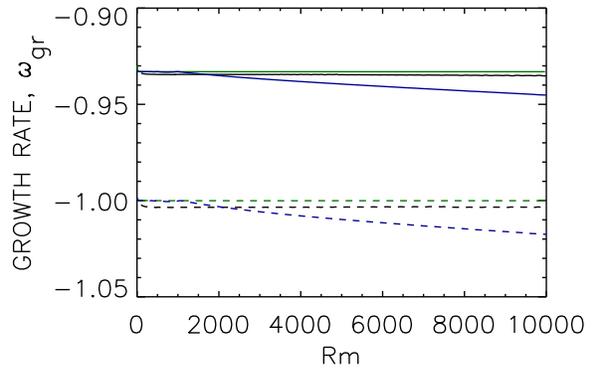}
 }
\caption{Growth rates   normalized with the
diffusion frequency for a nonaxisymmetric magnetic field with $m=1$.
Azimuth-dependent electric
conductivity  with  $\kappa=0.5$. $M=1$ (solid)  and $M=2$ (dashed).  Three different gaps between the cylinders:  $\rin=0.5$ (black), $\rin=0.9$ (green),  $\rin=0.95$ (blue).  $\mu=\rin^{1.5}$  (Keplerian rotation),
$k=1$. As in Fig. \ref{fig10} the drift rates vanish. Perfectly conducting
boundary conditions.
 }
\label{figM2}
\end{figure}

The upper panel of Fig. \ref{figM2} shows the  growth rates for Keplerian rotation which result from the approximations with $M=1$ and $M=2$ for a medium gap with $\rin=0.5$ and
 two   narrow gaps with $\rin=0.9$ and $\rin=0.95$. 
The number of equations strongly grows for growing $M$. For $M=1$  the solver works for twelve equations while for $M=2$  twenty equations are concerned. Nevertheless, the results are very similar, i.e.  the growth rates are always negative with almost the  same  numerical value of order unity. Hence, the modes decay with the diffusion time scale independent of the geometry of the tube and independent of the magnetic Reynolds number $\Rm$. We note that the diffusion time scale is formed with the outer cylinder radius $R_0$; the choice of $\rin$ does not a play a role for the decay rate of the  (nonaxisymmetric) mode. 

 In both approximations the decay rates behave very similarly. The main difference is a  slightly higher decay rate if $M=2$ which is {\em unity for all models} if $\Rm$ is small.  While for the wider gap the curves are strictly horizontal, there may be a little slope for the narrow gaps (see below). The lower panel of Fig. \ref{figM2} again demonstrates that the considered nonaxisymmetric field with $m=1$ does not drift in the azimuthal direction during its decay. The system always behaves very similar to the behavior of the mode $m=0$.  

\begin{figure}
  \centerline{
  \includegraphics[width=0.47\textwidth]{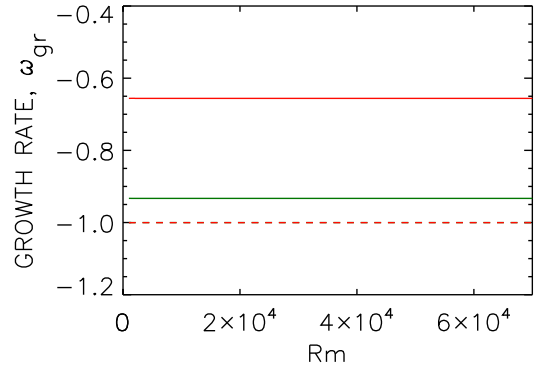}
  }
\caption{Similar to Fig. \ref{figM2} but for very large magnetic Reynolds numbers $\Rm$ and with $\kappa=0.5$ (green) and $\kappa=0.95$ (red). $M=1$ (solid), $M=2$ (dashed).  For $M=2$ the two dashed lines (red and green) are identical. $\rin=0.9$, $\mu=0.85$ (Keplerian rotation), 
$k=1$. As in Fig. \ref{fig10} the drift rates vanish in all cases. Perfectly conducting boundary conditions.}
\label{figRm5}
\end{figure}

The curves in Fig. \ref{figM2} do  not completely exclude the possibility  that for narrow gaps  the negative growth rates may change their sign for much  higher $\Rm$ and a dynamo could start to operate there. For the narrow  gap with $\rin=0.9$ we  thus evaluated with Keplerian rotation  the eigenfrequencies  also for very large magnetic Reynolds numbers and  for  the two  resistivity profiles with $\kappa=0.5$ and $\kappa=0.95$. $\kappa=0.5$ describes an azimuthal  peak-to-peak variation of the molecular conductivity by a factor of 3 and $\kappa=0.95$  by a factor of 39. The latter  model (thin gap, massive conductivity variations, large magnetic Reynolds numbers) may well approach the assumptions used by \cite{BW92} in flat geometry. But in  cylindrical geometry only  negative growth rates are provided for $\Rm\leq 10^5$,  independent of the actual value of $\kappa$ (Fig. \ref{figRm5}, upper panel). The nonaxisymmetric mode with $m=1$ just decays with the diffusion  time of the purely axisymmetric mode with $m=0$. The combination of differential rotation and non-uniform molecular conductivity in our calculations does not lead to positive growth rates, so that  a dynamo does  not exist here. Again the azimuthal drift of the magnetic pattern disappears as it does for the decay of the $m=0$ mode  shown in the lower panel of Fig. \ref{refer}.

We also briefly checked  the action of an  axial flow $U(R)$ --
independent of $z$ -- as a function of the radius for the induction  equations (\ref{g5}) and (\ref{g6}).  
\begin{figure}
  \centerline{
  \includegraphics[width=0.47\textwidth]{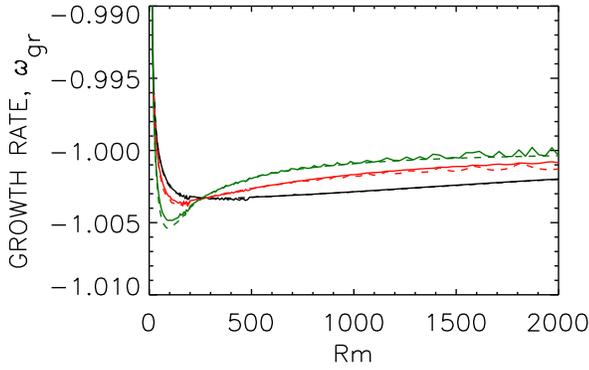}}
\caption{Growth rates in units of the
diffusion rate for various values of the axial flow $-0.5 \leq {\hat U}\leq 0.5$ 
for $k=1$ and
azimuth-dependent electric
conductivity. Solid lines: ${\hat U}>0$, dashed lines: ${\hat U}<0$. Green: $|{\hat U}|=0.5 $, red: $|{\hat U}|=0.25$, black: $|{\hat U}|=0.1 $. $\rin=0.5$, $\mu=0.35$ (Keplerian rotation). $\kappa=0.5$,
$M=2$. Perfectly conducting
boundaries.
 }
\label{figaxial}
\end{figure}
It is normalized with the azimuthal flow speed $\Om_{\rm in} R_0$, i.e.
$
U(R)= {u_z}/{\Om_{\rm in} R_0}$
and its radial profile is simply put to
\beg
U(R)= \hat U \sin \frac{2(R-\rin)\pi}{1-\rin}.
\label{axial}
\ende
Figure  \ref{figaxial} demonstrates the irrelevance of such an  axial flow (of both circulation regimes) for the dynamo mechanism, which is not surprising.  Positive results can only be expected after inclusion of a radial flow but then we would confirm --  or not -- the results of \cite{DJ89} for  the ability of combined systems of differential rotation and meridional circulations to solve the dynamo equation (\ref{mhd0}) with positive growth rates.
\section{Axial conductivity variations}
It remains to probe  axial profiles of the molecular resistivity to support dynamo action. 
This problem does not cover the transformation of the dynamo of \cite{BW92} to cylindrical or spherical geometry as they also denied the operation of such a dynamo in their model. Inspecting the equation system one finds a more complex instability problem. The last term of Eq. (\ref{mhd06}) for $\eta=\eta(z)$ provides a coupling of the equations via $B_z$. If ever, the radial magnetic field is exclusively originated by the axial field rather than by the azimuthal field as it is the case for $\eta=\eta(\phi)$. The coupling is thus only weak for dynamos with weak axial fields as it is usually realized  for dynamos with differential rotation. As nevertheless such a dynamo cannot be excluded new calculations must provide the consequences of profiles such as $\eta=\eta(z)$, see \cite{MG21}.

The axial profile of the resistivity may be modeled by
\beg
\eta=\eta_0(1-\zeta \cos z),
\label{axialcon}
\ende
where the free parameter $\zeta$ describes the amplitude of the axial resistivity variations. $\zeta=1$ must naturally be excluded. The wave number of the resistivity variations with (\ref{axialcon}) is thus unity.
The coefficients similar to  (\ref{kappas}) are 
\beg
\kappa_n=\frac{\i \zeta}{2\pi} \int^{2\pi}_0\frac{\sin z\  e^{\i n z}}{1-\zeta \cos z} \d z, \ \nonumber\\ 
\bar\kappa_n=\frac{ 1}{2\pi} \int^{2\pi}_0\frac{ e^{\i n z}}{1-\zeta \cos z}\d z.
\ende
Again  $\kappa_n$ and $\bar\kappa_n$ are real with $\kappa_{-n}=-\kappa_n$ and 
$\bar\kappa_{-n}=\kappa_n$.
The definitions
\beg
\XA_k=\frac{\d A_k}{\d R} + \frac{A_k}{R},\ \ \ \ \ \ \ \ \ \ \ \ \ \ \  \XB_k=\frac{\d B_k}{\d R} + \frac{B_k}{R}
\label{g32}
\ende
are also  parallel to  the above notation.
The resulting equations are
\beg
\frac{\d \XA_k}{\d R}  &-&( k^2+\frac{m^2}{R^2}) A_k  -\frac{2\i m}{R^2}B_k-\nonumber\\
&-&
\i {\Rm} (\omega+m\Om(R))\sum_{\ell=k^*-K}^{k^*+K} \bar\kappa_{\ell-k}  A_\ell\nonumber\\
&=&\sum_{\ell=k^*-K}^{k^*+K} \frac{\kappa_{\ell-k}}{\ell}  \left( \frac{\d \XA_\ell}{\d R}+\frac{\i m \XB_\ell}{R}   -\frac{2\i m B_\ell}{R^2}-\ell^2 A_\ell \right),\nonumber
\label{g2}
\ende
and
\beg
\frac{\d \XB_k}{\d R} & -&(k^2+\frac{m^2}{R^2}) B_k+\frac{2\i m}{R^2}A_k-\nonumber\\
&-&
  {\i \Rm}(\omega+m\Om(R))\sum_{\ell=k^*-K}^{k^*+K} \bar\kappa_{\ell-k}B_\ell-\nonumber\\
&-&
 2{\Rm}\ \frac{b_\Om}{ R^2} \sum_{\ell=k^*-K}^{k^*+K}\bar\kappa_{\ell-k}A_\ell\nonumber\\
&=&\sum_{\ell=k^*-K}^{k^*+K} \frac{\kappa_{\ell-k}}{\ell}\left( \frac{\i m \XA_\ell}{R}   -\frac{ m^2 B_\ell}{R^2}-\ell^2 B_\ell\right),
\label{g3}
\ende
formulated for all $k$ with $k^*-K\leq k\leq k^*+K$. Although the $k$'s denote  wave numbers they are here considered  as real integers\footnote{We note that here  the asterisk does not mean ``complex conjugate''.}.  The system   contains $8K+4$ equations with $k^*$  as the central wave number. Because of the Cowling theorem  only the stability of nonaxisymmetric fields is probed, hence $m=1$ is fixed.

\begin{figure}
  \includegraphics[width=0.47\textwidth]{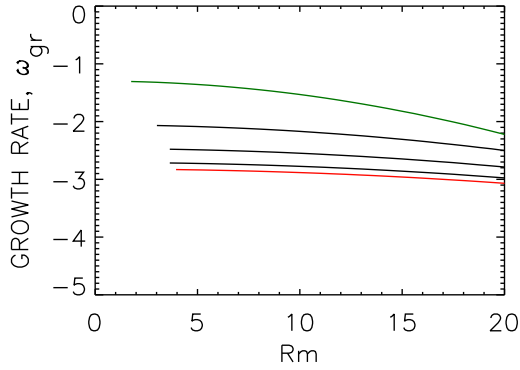}
\caption{Growth rates for containers with the axial conductivity profile (\ref{axialcon}) for slow rotation.    It is  $k^*=2$
 $\rin=0.5$, $\mu=0.35$ (Keplerian rotation), $m=1$.  The $\zeta$ varies from 0.9 (green) to $\zeta= 0.1$ (red) .
 Perfectly conducting boundary conditions.}
\label{figaxial4}
\end{figure}

We start with $k^*=2$  and  $K=1$ so that a modal system with wave numbers $k=1$, $k=2$ and $k=3$ is considered.  The wave number $k=1$ of the resistivity variation  (\ref{axialcon}) is part of the spectrum. The remaining modes $k=2$ and $k=3$ are higher modes. Indeed,  Fig. \ref{figaxial4} demonstrates  that for small $\zeta$ and slow rotation the decay rates approach that for $k=1$ and {\em uniform} resistivity shown by Fig. \ref{refer} for the case of $m=1$. Faster rotation seems to amplify the decay of the magnetic field.

\begin{figure}
  \centerline{
 \vbox{
  \includegraphics[width=0.47\textwidth]{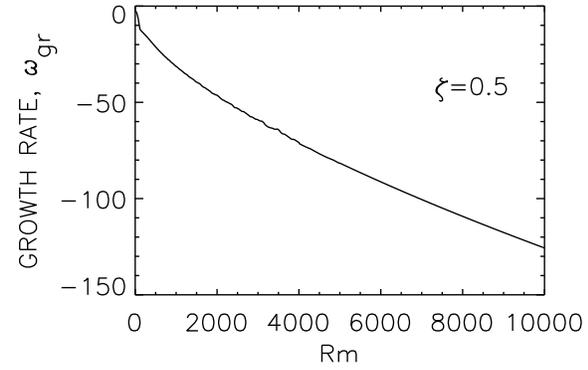}
  \includegraphics[width=0.47\textwidth]{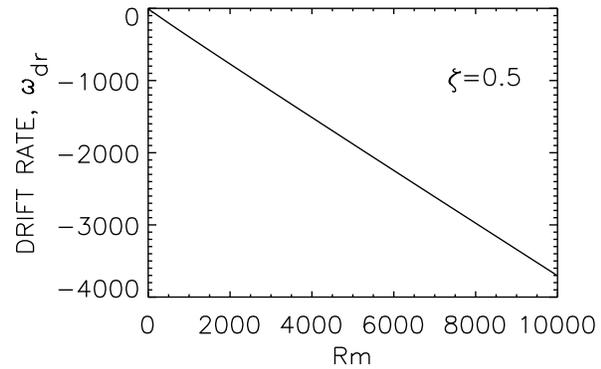}}
  }
\caption{Growth rate (top) and drift rate (bottom) for containers with the periodic 
axial conductivity profile (\ref{axialcon}) with variation of $\Rm$ and for $m=1$.   The axial modulation of the resistivity  is fixed to 
$\zeta=0.5$. It is $k^*=2$.
 $\rin=0.5$, $\mu=0.35$ (Keplerian rotation), 
 Perfectly conducting boundary conditions.}
\label{figaxial1}
\end{figure}

The overall result of the calculations is that we do not find   modes for any magnetic Reynolds number    and/or for any value of $\zeta$ as unstable, i.e.  all growth rates become negative. Figure \ref{figaxial1} demonstrates for nonaxisymmetric fields with $m=1$, axial resistivity distribution and much higher Reynolds numbers that  both the physical growth rate and the physical drift rate  grow with growing rotation rate,  similar to  the decay mode with $m=1$ and $\zeta=0$ in Fig. \ref{refer} (red lines). Differential rotation obviously destabilizes the magnetic fields, i.e. they  decay faster.  On the other hand,  the axial 
non-uniformity of the molecular conductivity slightly {\em stabilizes} the magnetic field in comparison to the same configuration with uniform conductivity.

\begin{figure}
  \includegraphics[width=0.47\textwidth]{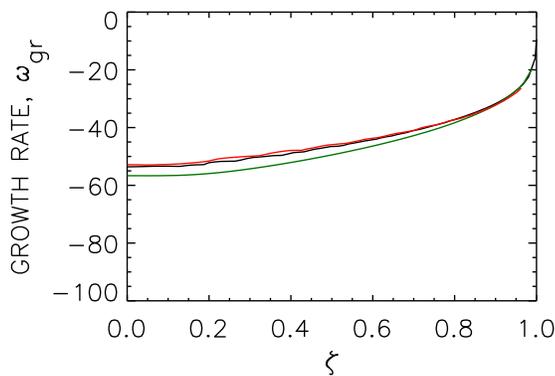}
\caption{Growth rates for containers with the axial conductivity profile (\ref{axialcon}) for variation of $\zeta$.   The magnetic Reynolds number is fixed to 
$\Rm=2000$. It is $ k^*=1.5$ (red), $k^*=2$ (black) and $k^*=3$ (green).
 $\rin=0.5$, $\mu=0.35$ (Keplerian rotation), $m=1$. We note that possible solutions for the exotic case  $\zeta=1$ are excluded. 
 Perfectly conducting boundary conditions.}
\label{figaxial2}
\end{figure}

Another example, Fig. \ref{figaxial2},  contains the growth rates and the drift rates for a fixed  high value $\Rm=2000$ for the three central wavenumbers $k^*=1.5$, $k^*=2$ and $k^*=3$.
For $\zeta=0$ and $k^*=2$ both the growth rate and the drift rate are identical to the numbers given   in Fig. \ref{refer} for $\Rm=2000$ while for $\zeta=0.5$ the values for $\Rm=2000$ in the Figs. \ref{figaxial1} and \ref{figaxial2}  coincide. For growing $\zeta$ the absolute value of the growth rate sinks while the drift rate remains almost constant. The differences of these values for different wave numbers $k^*$ are very small. Obviously, the fastest decay of the magnetic field happens for $\zeta=0$, i.e. if the resistivity in the cylinder is uniform.  

  {Dynamos with differential rotation without meridional circulation   have never been found in this study even not for models  with  non-uniform  resistivity profiles}.

\section{Conclusion}
We have shown by use of both spherical and cylindrical models that any   dependence of the molecular resistivity $\eta $ on the position $\vec{x}$ does not soften the toroidal-velocity antidynamo theorem of \cite{E46}. Our laminar velocity fields do not possess 
 radial components, and  thus we do not  find  any  dynamo self-excitation. Under the exclusive influence of various rotation laws the nonaxisymmetric $m=1$ mode {\em decays} for all the considered   $\eta$-profiles and gap widths up to magnetic Reynolds numbers of $10^5$. The decay time runs  the diffusion time scale for axisymmetric disturbances and with the rotation time for nonaxisymmetric disturbances (and fast rotation) almost independent of the amplitude of the azimuthal variation of the the $\eta$-profile and also independent of the differential rotation laws.

These  findings should  be  of relevance for the idea that a hot exoplanet  develops strong differences of the  electric conductivity at their day-side and  night-side with possible  potential to form a magnetic dynamo system  \citep{RM17}. Based on our results this would  only be possible if the  antidynamo theorem is overcome by inclusion of a radial flow component, but such complex  flows can already generate nonaxisymmetric magnetic fields even without  non-uniform molecular conductivity \citep{DJ89}. For the operation of a laminar dynamo the  radial flow component  cannot successfully  be replaced by gradients of the electric conductivity. The  idea of Elsasser about resistivity variations  is confirmed that  ``it does not seem   plausible that  this variation  introduces phenomena
that modify the theory in a fundamental
way''.

\noindent
{\bf Acknowledgment} Axel Brandenburg (Stockholm), Rainer Hollerbach (Leeds) and
 Johannes Wicht (G\"ottingen) are cordially acknowledged for stimulating discussions.

\bibliographystyle{mn2e}
\bibliography{superamri}
\end{document}